%% file: ms.tex
\documentclass[runningheads]{llncs}
\usepackage[T1]{fontenc}
\usepackage{graphicx}
\usepackage{amssymb}
\usepackage{booktabs} 
\usepackage{enumitem}
\usepackage{amsmath}
\usepackage{subcaption}
\usepackage{verbatim} 
\usepackage[export]{adjustbox}
\usepackage{array}
\usepackage{soul,color} %%%{hl}
\usepackage{multirow}
\usepackage{multicol}
\usepackage{stackengine}
\usepackage{bm}
\usepackage{color} % If not already included
\definecolor{RemovedText}{rgb}{1,0,0} % Red color

\usepackage[normalem]{ulem}
\usepackage{color}
\definecolor{AddedText}{rgb}{0,0,1} % Blue color for additions
\definecolor{RemovedText}{rgb}{1,0,0} % Red color for removals

\usepackage[capitalize]{cleveref}

\input{commands.tex}

\newcommand\MyIm[1]{%
	\includegraphics[width=1.6cm,height=1.6cm]{#1}
}
\newcommand\MyImn[1]{%
	\includegraphics[width=1.3cm,height=1.3cm]{#1}
}
\newcommand\MyImm[1]{%
	\includegraphics[width=1.4cm,height=1.4cm]{#1}
}
\newcommand\MyImmm[1]{%
	\includegraphics[width=1.5cm,height=1.5cm]{#1}
}
\newcommand\MyTiny[1]{\fontsize{1}{1}\selectfont{#1}}
\newcolumntype{E}{ >{\centering\arraybackslash} m{1.6cm}}
\newcolumntype{C}{ >{\centering\arraybackslash} m{1.4cm}}
\newcolumntype{D}{ >{\centering\arraybackslash} m{1.5cm}}
\newcolumntype{F}{ >{\centering\arraybackslash} m{1.3cm}}

\begin{document}
\title{Reconstructing Interpretable Features in Computational Super-Resolution microscopy via Regularized Latent Search}
\titlerunning{Reconstructing Interpretable Features via Regularized Latent SearchY}
\author{Marzieh Gheisari \and
Auguste Genovesio}
\institute{Institut de Biologie de l'ENS, Ecole Normale Supérieure, 46 rue d'Ulm, PSL Research University, 75005 Paris}
\maketitle         
\begin{abstract}
Supervised deep learning approaches can artificially increase the resolution of microscopy images by learning a mapping between two image 
resolutions
or modalities. However, such methods often require a large set of hard-to-get low-res/high-res image pairs and produce synthetic images with a moderate increase in resolution. Conversely, recent methods based on GAN latent search offered a drastic increase in resolution without the need of paired images. However, they offer limited
reconstruction
of the high-resolution image
interpretable features.
Here, we propose a robust super-resolution method based on regularized latent search~(RLS) that offers an actionable balance between fidelity to the ground-truth and realism of the recovered image given a distribution prior. The latter allows to split the analysis of a low-resolution image into a computational super-resolution task performed by deep learning followed by a quantification task performed by a handcrafted algorithm
and based on interpretable biological features. This two-step process holds potential for various applications such as diagnostics on mobile devices, where the main aim is not to recover the high-resolution details of a specific sample but rather to obtain high-resolution images that preserve explainable and quantifiable differences between conditions.
\end{abstract}

\section{Introduction}
Various deep learning models have shown excellent performance in the single-image super-resolution (SISR) task which aims to restore a high-resolution (HR) image from its low-resolution (LR) counterpart. Deep-learning SISR models have been applied few years ago to enhance the resolution of microscopy images~\cite{weigert2018content,wang2019deep,ouyang2018deep}.
More recently, studies have explored image-to-image translation models, that are trained to learn a parameterized function between two different
image resolutions or modalities. These supervised approaches require a large number of paired images and rely on generative models that output artificial images~\cite{hoffman2021promise}. These artificial images were accepted by the community as real mainly because they were measured as approaching real image instances. However, for the latter to hold, these approaches only offered a moderate increase in resolution of up to $4$x.

Importantly, learning a parameterized mapping from a low- to a high-resolution image is an ill-posed problem: a single low-resolution image corresponds to infinitely many highly resolved ones. Therefore, the super-resolution task cannot only consist of maximizing the fidelity of the recovered high-resolution image; it must be further constrained to be well-posed. In this work, we propose to constrain this task by imposing
that the reconstructed super-resolution image be realistic, \ie, belongs to a given image distribution,
a principle that is also explored by utilizing gradient distribution prior in the context of biomedical images~\cite{gong2015natural}.
In short, on the one hand, fidelity refers to recovering a super-resolution image that, once downgraded, is close to the original image. On the other hand, realism refers to keeping the image within a given image domain. 

In this paper, we aim to better define the super-resolution task by enforcing the solution to be a trade-off between fidelity
to the original low-resolution image and realism, that includes the preservation of biologically relevant content, which we describe as ``interpretable features''.
Assessing these features is essential for accurate phenotypic interpretation and discrimination. By guiding the super-resolution process towards this end, we
% }
generate images that are both faithful to the sample and biologically interpretable and quantifiable. In this way, we anticipate super-resolution in biology could exploit more than recovering only the details of a given sample, but benefit from features measured over a set of recovered images.

Our approach, Regularized Latent Search~(RLS), consists of a regularized search in the latent space of a pre-trained generative model for high-resolution images. By doing so, we both suppress the need for paired images and make the problem well-posed as we search for the closest image a generator can produce that, when down-scaled, matches the low-resolution image input.
Moreover, as we keep the super-resolved image in the original domain, instead of producing a moderate increase in resolution, we propose to push further the synthesis: we anticipate that creating very highly resolved (up to $32$x) but controlled artificial images could be of great interest for applications such as diagnostic. This is because what is at stake, in this case, is the preservation of measurable %\added{
and interpretable features
of microscopy images, not the absolute matching with real image samples.

Similar to image-to-image translation methods like CycleGAN\cite{zhu2017unpaired}, which operate without the need for paired images, our method differentiates itself in its approach to super-resolution. It is specifically designed to address the ill-posed nature of super-resolution by utilizing a GAN’s latent space to identify plausible HR images for any given LR input. Once the model is trained, it enables the super-resolution of any LR image with just an adjustment in the degradation function during inference. Unlike CycleGAN, which is deterministic and tailored for specific domain translations requiring individual models for each task, our method offers versatility and is suitable for a broad spectrum of super-resolution tasks. It eliminates the need to learn domain-specific features and is intended for widespread application in super-resolution, capable of being trained on a single dataset to enhance any LR image.
 \section{Related Work and Background}
 \subsection{Super-resolution of microscopy images}
While optical super-resolution techniques such as STED~\cite{hell1994breaking} and PALM~\cite{betzig2006imaging}
can break the diffraction limit, they are limited by the need for specialized equipment and complex sample preparation.
Computational super-resolution methods on the other hand, while they rely on existing training data, can represent a cheap way to enhance the resolution of images acquired with conventional microscopes.
Several approaches were developed to address the problem of computational super-resolution of microscopy images. Deep-STORM~\cite{nehme2018deep} uses an encoder-decoder network to localize emitters in super-resolved images. Content-aware image restoration~\cite{weigert2018content} uses a U-Net architecture and is trained with low signal-to-noise ratio (input) and high signal-to-noise ratio (target) image pairs. These approaches produced interesting results but recovered images often lacking high-frequency details due to the MSE loss. To balance this issue, other approaches such as ANNA-PALM~\cite{ouyang2018deep} are based on generative adversarial network (GAN). The authors trained a U-Net, and in contrast with Deep-STORM, a combination of pixel-wise reconstruction loss and adversarial losses is used to obtain image reconstructions of better quality. Using a similar network architecture and training loss, Wang et al.~\cite{wang2019deep} achieved super-resolution in fluorescence microscopy across different modalities. Overall, using an adversarial loss results in output images that are sharper and of better perceptual quality and these methods demonstrated the potential of deep learning to improve the spatial resolution of fluorescence microscopy images. While the deep learning architectures and the applications of these methods differ, they all require hard-to-get paired image data for training and offer a moderate increase in resolution of up to $4$x.

It is essential to note that computational super-resolution techniques effectively enhance the resolution of images by optimizing the use of available data in the original images. However, these methods are constrained by the existing information and cannot add details beyond what the original optical systems could capture. 

\subsection{Style-based generative models}
\label{sec:stylebased}
StyleGAN models~\cite{karras2019style,karras2020analyzing} are well-known for their ability to generate highly realistic images.
The StyleGAN architecture consists of two sub-networks: a mapping network denoted by $\G_m:\real^\dim \rightarrow \real^\dim$, and a synthesis network consisting of $\layer$ layers denoted by $\G_s:\real^{\layer\times\dim} \rightarrow \real^{\n}$. Here, $\dim$ represents the dimensionality of the latent space. The mapping network takes a sample $\z\in\real^\dim$ from a standard normal distribution and maps it to a vector $\w\in\wspace$, where $\wspace$ denotes the intermediate latent space. StyleGAN2 introduced path length penalty, which encourages a fixed-size step in $\w$ to result in a fixed-magnitude change in the image.
The regularization term is computed using the Jacobian determinant, and penalizing changes in it promotes the generation of smoother and more realistic images.
$\layer$ copies of the $\dim$-dimensional vector $\w$ are fed to the $\layer$-layer synthesis network $\G_s$, with each copy
representing the input to the corresponding layer of $\G_s$. The $\G_s$ network controls the level of detail in the generated image
at each layer.
Individual modification of these $\layer$ layers, by adjusting the latent vector copy of $\w$ for each layer, extends the latent space into$\wspace^+$.
This extended latent space enhances the model’s capability for accurate image reconstruction, which is vital for super-resolution tasks,
offering more nuanced control over the image generation process ~\cite{abdal2019image2stylegan,wulff2020improving}.

 \subsection{GAN-based high-resolution image reconstruction}
 The problem of obtaining a super-resolution (SR) image of dimension $n$ from a low resolved (LR) image of dimension $m$ is ill-posed as, for a non-invertible forward operator $\D$ with $m<n$, there are infinitely many high resolution (HR)  images that match a given LR image. Thus the reconstruction procedure must be further constrained by prior information to better define the objective and lead to a stable solution.
 One such prior consisted of considering the reconstructed HR image to be part of a given domain. First, the distribution of HR images is learned in an unsupervised fashion, thanks to a GAN, then the latent space of this trained GAN is searched to find a latent vector producing an HR image that, once down-scaled, is the closest to the LR image input.
 GAN-prior-based images reconstruction 
 was first introduced by Bora \etal~\cite{bora2017compressed} and further improved using StyleGAN~\cite{karras2019style,karras2020analyzing} by Menon \etal~ in PULSE~\cite{menon2020pulse} by constraining the search to remain on the image manifold. To this end, PULSE and two other studies~\cite{zhu2020improved,wulff2020improving} use an invertible transformation of the intermediate latent space $\wspace^+$ which includes a leaky rectified linear unit (ReLU)~\cite{goodfellow2016deep} followed by an affine whitening transformation, so that transformed latent vectors approximately followed the standard Gaussian distribution $\normal(\zero, \eye_\dim)$. Sampled vectors are then constrained to lie around a hypersphere with radius $\sqrt{\dim}$ hypothesizing that most of the mass of a high-dimensional Gaussian distribution is located at or near $\sqrt{\dim} \sphere^{\dim-1}$, where $\sphere^{\dim-1}$ is the $\dim$-dimensional unit hypersphere. Constraining samples to lie in a dense area of the StyleGAN style distribution resulted in increased realism of the generated images.

 Although the above approach showed major improvements over previous work, it also presents important caveats that in practice led to image artifacts.
Here we show that transforming the intermediate latent space in this way does not lead to an accurate standard Gaussian distribution,
and so prevents proper regularization based on this hypothesis. Moreover, as the search is strictly limited to the spherical
surface $\sphere^{\dim-1}$, this limitation may prevent the search from reaching the closest reconstruction of the high-resolution image.

In this work, we regularize the search in the latent space for a latent code located in ``healthy'' regions of the latent space. In this way, the system is constrained to produce images that belong to the original image domain StyleGAN was trained on. To do so, we take advantage of normalizing flow to gaussianize the latent style sample distribution which leads to a much closer standard Gaussian distribution.
We then use this revertible transformation to regularize the search in $\wspace^+$ such that it remains in a high-density area of the style vector distribution. We then show experimentally that the latter produces reconstructed images that are not only realistic but also more faithful to the original HR image.

\section{Method}
\label{sec:RLS}

\subsection{Super-resolution by Regularized Latent Search}
Super-resolution aims to reconstruct an unknown high resolution-image $\x\in\real^{\n\times \n}$ from a low-resolution image $\y\in\real^{m\times m}$, which is related to the HR image by a down-scaling process described by $\y=\D(\x)+\deltaa$, where $\D:\real^{n\times n}\rightarrow\real^{m\times m}$ is a non-invertible down-scaling forward operator and $\deltaa$ is an independent noise with distribution $\Pdelta$.
We can formulate this task in terms of Maximum A Posteriori (MAP) estimation~\cite{marinescu2021bayesian}. Given an LR image $\y$, our goal is to recover the HR image $\x$ as the MAP estimate of the conditional distribution $\Pg(\x|\y)$:
\begin{equation}
\label{equation mapestimation}
\begin{aligned}
\arg\max_{\x}\log \Pg(\x|\y) = \arg\max_{\x} [\log \Pdelta(\y-\D(\x)) + \log \Pg(\x)].
%& \arg\max_{\x}\log \Pg(\x|\y) &=& \arg\max_{\x} [\log p(\y|\x) + \log \Pg(\x) + \log p(\y)]\\
%& \qquad &=& \arg\max_{\x} [\log \Pdelta(\y-\D(\x)) + \log \Pg(\x)]
\end{aligned}
\end{equation} 
The first term is the likelihood term describing the image degradation process $\D$ and the second is the image prior, describing the manifold of real HR images.

\subsubsection{Image prior} Let $\G_s$ be the synthesis network of a StyleGAN~\cite{karras2020analyzing} pretrained on the considered image domain. $\G_s$ takes as input $\w$, produced by the mapping network, and outputs an image.
The image prior $\log \Pg(\x)$ can be expressed with respect to the latent variables $\w$:
\begin{equation}
\label{equation changeofvarivale}
\log \Pg(\G_s(\w)) = \log \Pw(\w) + \log |\det J_{\G_s^{-1}} (\w)|.
\end{equation}
The second term can be dropped as the path length penalty in StyleGAN2 implies that the Jacobian determinant is constant for all $\w$. The first term $\Pw(\w)$ is the image prior we define on $\w \in \wspace^+$ by:
\begin{equation}
\label{equation imageprior}
\log \Pw(\w) =  \lambda_w \prior_{w} + \lambda_c \prior_{cross} %+ \lambda_g \prior_{gaussian} 
\end{equation}
where:
%\noindent
\begin{itemize}[leftmargin=6mm, label={\tiny$\bullet$}]
	\item $\prior_{w}$ is a prior that keeps $\w$ in the area of high density in $\wspace^+$: $\prior_{w} = \frac{1}{\layer}\sum_{i=1}^{\layer} {\log \Pf(\w_i)}$
	, where $\Pf(\w)$ is estimated by a normalizing flow model $\F$ explained in further detail below.
	\item $\prior_{cross}$ is a pairwise euclidean distance prior on $\w  = [\w_1,\dots, \w_\layer]\in \wspace^+$ that ensures $\w \in \wspace^+$ remains close to the trained manifold in $\wspace$:~ $\prior_{cross} = -\sum_{i=1}^{\layer-1}\sum_{j=i+1}^{\layer}{||\w_i-\w_j||^2_2}$
	\item $\lambda_w$ and $\lambda_c$ are hyperparameters that control the relative importance of the two priors.
\end{itemize}

\subsubsection{Normalizing Flow}
Using a sequence of invertible mappings, a Normalizing Flow $\F:\real^\dim\rightarrow\real^{\dim}$ is a transformation of an unknown complex distribution into a simple probability distribution that is easy to sample from and whose density is easy to evaluate such as standard Gaussian~\cite{kobyzev2020normalizing}.

Let $\z=\F(\w)$ with probability density function $p(\z)$. Using the change-of-variable formula, we can express the log-density of $\w$ by~\cite{papamakarios2021normalizing}:
\begin{equation}
\log \Pf(\w) = \log p(\z) + \log |\det J_{\F} (\w)|, \qquad \z=\F(\w)
\label{equation nf}
\end{equation}
where $J_\F(\w)$ is the Jacobian of $\F$ evaluated at $\w$. In practice the Jacobian determinant in \cref{{equation nf}} should be easy to compute so that the density $\Pf(\w)$ can be evaluated. Furthermore, as a generative model, the invertibility of $\F$ allows new samples $\w = \F^{-1}(z)$ to be drawn through sampling from the base distribution. In the literature, several flow models were proposed, such as Real Non-volume Preserving Flow
~\cite{dinh2016density} and Masked Auto-regressive Flow (MAF)~\cite{papamakarios2017masked}.

\subsubsection{Optimization process}
In the likelihood term presented in \cref{equation mapestimation}, we assume that the noise follows a Laplace distribution, that is $\deltaa\sim Laplace(0, \lambda_l I)$. This assumption simplifies the log-density of $\deltaa$ to: $\log \Pdelta (\deltaa)=-{||\deltaa||_1}-C$, where $C$ is a constant. Thus the optimization problem in \cref{equation mapestimation} is effectively transformed into an optimization over $\w$, resulting in the final objective function:
\begin{equation}
\label{equation loss}
\begin{aligned}
\hat{\w} = \arg\min_{\w} ||\y - \D(\G_s(\w))||_1 - \log \Pw(\w)
\end{aligned}
\end{equation}

The selection of the Laplace distribution for noise modeling is favored due to its simple form, which simplifies the log-density function and makes the loss function easier to optimize. Moreover, the $L_1$ norm, which arises in the log-density of the Laplace distribution, offers greater robustness to outliers than the $L_2$ norm. This characteristic is particularly beneficial in the context of super-resolution, where the ability to handle irregular noise or artifacts in low-resolution images is crucial, especially given the challenges commonly associated with microscopy data.

\begin{figure}
\centering
\includegraphics[width=0.65\textwidth]{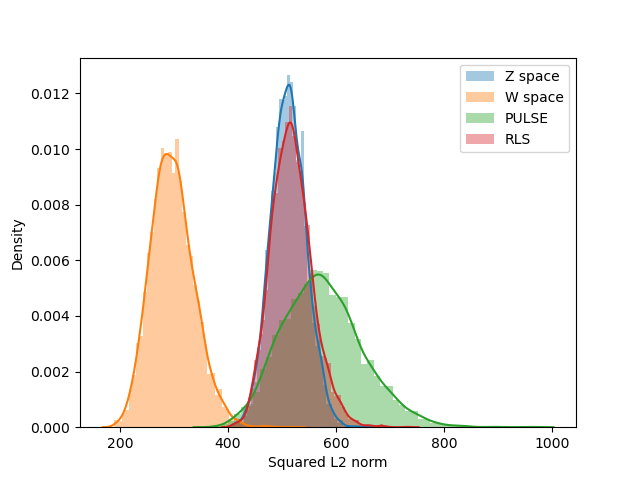}
\caption{
Distribution analysis of squared $L_2$ norms demonstrating the gaussianization of latent style vectors.
The graph compares the density of squared norms from the original $\zspace$ space (blue), untransformed $\wspace$ space (orange),
and the distributions resulting from the PULSE (green) and our method (red).}
% }
\label{fig:gaussian}
\end{figure}

\subsection{Evaluation of the Gaussianization process}
To assess the quality of the gaussianization process, we sampled 5000 vectors $\z \sim \normal (\zero, \eye_\dim)$.
These vectors were then transformed into style vectors $\w=\G_m(\z)$ using the mapping network $\G_m$ of StyleGAN.
We then gaussianized the distribution of $\wspace$ using alternatively PULSE and our method.
The quality of the gaussianization process was evaluated by computing the squared norm for all of these vectors
within the transformed distribution (see \cref{fig:gaussian}).
Ideally, the squared norm of the standard Gaussian distribution $\zspace$ should follow a chi-squared distribution
\ie $||\z_n||^2_2 \sim \chi^2_\dim$ and thus forms a narrow distribution centered around $\dim=512$, the dimensionality of $\wspace$. We can see that this is the case for $\zspace$ which is Gaussian. However, the $\wspace$ does not follow this pattern, which is inconsistent with the prior assumption held by BRGM~\cite{marinescu2021bayesian}. Furthermore, while the PULSE method results in a broader distribution, our method more accurately narrows the squared norm distribution to match the expected chi-squared distribution.

\section{Experiments}
\subsection{Experimental settings}
\label{experimental_setting}
A critical aspect of our study was the necessity to work with high downscaling factors, such as 16x and 32x, which are
central to our application scenario.
These factors are not commonly used in the literature, to the best of our knowledge there are no experimental datasets available with such high downscaling factors.
Most existing datasets typically focus on moderate downscaling factors of 2x or 4x, which is
not aligned with the needs of our study, where we aim to address more extreme cases of resolution enhancement.
Therefore, we had to generate our own low-resolution images.

As for implementation details, we began by training a StyleGAN2-ADA~\cite{karras2020training} model on a subset of the
BBBC021 dataset~\cite{caicedo2018weakly,ljosa2013comparison} which comprises cells treated with drugs acquired following the cell painting assay~\cite{caie2010high}. That is, cells were fluorescently stained with markers for F-actin, B-tubulin, and DNA, as described in~\cite{caie2010high}.
The dataset comprises wide-field epi-fluorescence images, captured using the automated ImageXpress imaging platform. 
From this dataset, we extracted images of size $128\times128$ centered around each cell nucleus and used 400 images per compound treatment (approximately 28,000 in total) to train StyleGAN2-ADA~\cite{karras2020training}. We then proceeded to generate 100000 random samples  $\z \sim \normal (\zero, \eye_\dim)$ and used their associated style vector $\w=\G_m(\z)$ to train the normalizing flow model. We opted for MAF, as it tends to perform better than RealNVP for density estimation tasks. Our normalizing flow model comprised five flow blocks with all hidden dimensions set to 1024. For super resolving the images using our regularized latent search algorithm, we employed an Adam optimizer over 200 iterations with a learning rate of 0.5 and initialized the search using the mean of 10,000 randomly generated latent vectors.

To fine-tune the regularization parameters $\lambda_w$ and $\lambda_c$, we empirically evaluated a spectrum of values, eventually setting them to 5e-5 and 0.01, respectively. This process revealed that the algorithm's performance remained relatively stable across a broad range of these parameters, indicating a lack of sensitivity as long as the values fell within a specific boundary. Precisely, we observed optimal performance when $\lambda_w$ was between 1e-6 and 5e-4, and $\lambda_c$ ranged from 0.005 to 0.05. These findings suggest that while the exact values of $\lambda_w$ and $\lambda_c$ are flexible, maintaining them within these determined ranges ensures the algorithm functions effectively.

\setlength{\tabcolsep}{0pt}
\begin{figure*}[!h]
	\centering
	\begin{scriptsize}
		\begin{tabular}{CEEEEEE}
			& LR & Pix2Pix & PULSE & BRGM & RLS & GT\\
			\raisebox{7mm}{\rotatebox[origin=c]{90}{\MyTiny{\stackanchor{CytochalasinB}{30$\mu$M}}}} &
			\MyIm{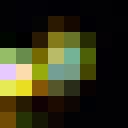} &
			\MyIm{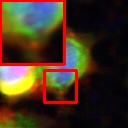} &
			\MyIm{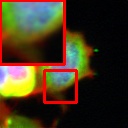} &
			\MyIm{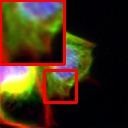} &
			\MyIm{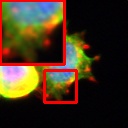} &
			\MyIm{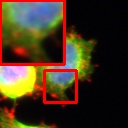}\\
   			\raisebox{7mm}{\rotatebox[origin=c]{90}{\MyTiny{\stackanchor{HerbimycinA}{10$\mu$M}}}} &
			\MyIm{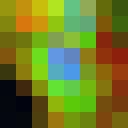} &
			\MyIm{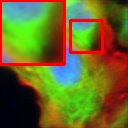} &
			\MyIm{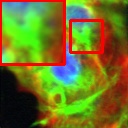} &
			\MyIm{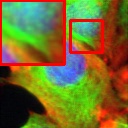} &
			\MyIm{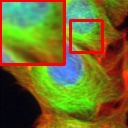} &
			\MyIm{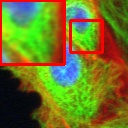}\\
			\raisebox{5mm}{\rotatebox[origin=c]{90}{\MyTiny{\stackanchor{Nocodazole}{3$\mu$M}}}} &
			\MyIm{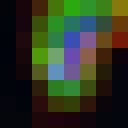} &
			\MyIm{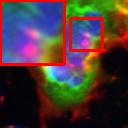} &
			\MyIm{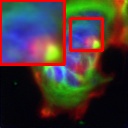} &
			\MyIm{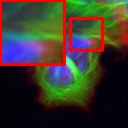} &
			\MyIm{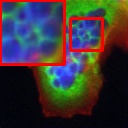} &
			\MyIm{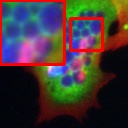}\\
			\raisebox{7mm}{\rotatebox[origin=c]{90}{\MyTiny{\stackanchor{CytochalasinB}{1$\mu$M}}}} &
			\MyIm{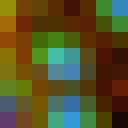} &
			\MyIm{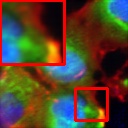} &
			\MyIm{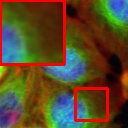} &
			\MyIm{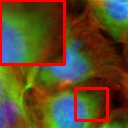} &
			\MyIm{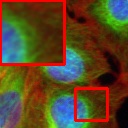} &
			\MyIm{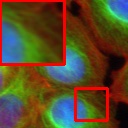}\\
			\raisebox{7mm}{\rotatebox[origin=c]{90}{\MyTiny{\stackanchor{DMSO}{}}}} &
			\MyIm{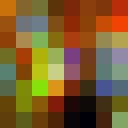} &
			\MyIm{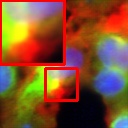} &
			\MyIm{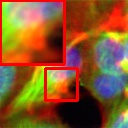} &
			\MyIm{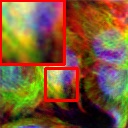} &
			\MyIm{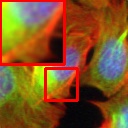} &
			\MyIm{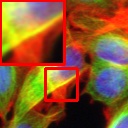}\\
		\end{tabular}
	\end{scriptsize}
	\caption{Qualitative comparison of super-resolution reconstructions on the BBBC021 dataset:
	visualizing the performance of RLS against baseline methods in reconstructing cellular structures
	and phenotypes under negative control (DMSO) and various treatment conditions at a 16x upscaling factor.
	}
	\label{fig:qualitative bbbc021}
\end{figure*}

For evaluation, we used 100 random samples from various compound treatments, ensuring samples were not included in the training of the generative network. We simulated degraded images from these high-resolution images using bicubic down-sampling, which is a common choice in the literature. We then compared the results of our algorithm with those of Pix2Pix and two state-of-the-art unsupervised image reconstruction methods based on StyleGAN inversion: PULSE~\cite{menon2020pulse} and BRGM~\cite{marinescu2021bayesian}. For the baseline methods, we used the same parameter settings reported by the original papers.

\subsection{Results}
\subsubsection{RLS recovers high-quality cell images}
We reconstructed images using RLS and compared them to images generated with baseline methods. The results of this comparison is presented in \cref{fig:qualitative bbbc021}.
At these high upscaling factors, due to the lack of proper regularization, the competing methods fail to produce reasonable cellular details and tend to produce images that can be accurately downscaled to the LR image at the cost of generating distorted cellular structures. In contrast, our approach produces highly realistic images for most examples and can reproduce cell phenotypes induced by compound treatment even if the fine-grain details are not similar. For instance, when reconstructing the images of cells treated with  Nocodazole, a known microtubule destabilizer, RLS captures the typical fragmented microtubule phenotype, resulting in a more accurate texture. Reconstructing images of cells at higher resolution obviously provides access to finer-grain measurable features.

To quantitatively assess this gain in performance, we evaluate the SR images using Frechet Inception Distance (FID)~\cite{heusel2017gans} and Kernel Inception Distance (KID)~\cite{binkowski2018demystifying} to measure the discrepancy between the real HR images and the reconstructed one. As expected, the scores of FID and KID listed in \cref{tab:quantitative} show that with both upscaling factors, our method significantly improves realism.
\setlength{\tabcolsep}{4pt}
\begin{table*}
	\centering	
	\begin{tabular}{@{}l||l|ll|lll@{}}
		\toprule
		Upscaling factor & Method & FID$\downarrow$ & KID$^{({\times10^3})}$$\downarrow$ & MSSIM$\uparrow$ & LPIPS$\downarrow$ & PSNR$\uparrow$\\
		\hline
		\multirow{5}{*}{32x} & Pix2Pix~\cite{lee2019enhancing} & 67.141 & 66.515 & 0.507 & 0.420 & 15.846\\
		& PULSE~\cite{menon2020pulse} & \textit{13.629} & \textit{9.056} & \textit{0.526} & \textit{0.378} & \textit{16.469}\\
		& BRGM~\cite{marinescu2021bayesian} & 15.862 & 13.293 & 0.400 & 0.475 & 14.207\\
  & RLS %0.0005,0,0.1
  &  \textbf{7.571} & \textbf{3.209} & \textbf{0.528} & \textbf{0.343} & \textbf{16.536}\\
\hline
		
		\multirow{5}{*}{16x} & Pix2Pix~\cite{lee2019enhancing} & 45.503 & 41.700 & \textit{0.708} & 0.274 & \textit{19.983}\\
		& PULSE~\cite{menon2020pulse} & \textit{17.354} & \textit{14.937} & \textbf{0.723} & \textbf{0.244} & \textbf{20.806}\\
		& BRGM~\cite{marinescu2021bayesian} & 19.862 & 17.254 & 0.608 & 0.301 & 17.914\\
		& RLS & \textbf{6.983} & \textbf{3.347} & 0.660 & \textit{0.262} & 19.177\\
		\bottomrule
	\end{tabular}
	\caption{
Quantitative evaluation of RLS and baseline methods for super-resolution on the BBBC021 dataset
at 32x and 16x upscaling factors. The \textbf{best} and the \textit{second-best} are highlighted in bold and italic respectively.}
\label{tab:quantitative}
 \vspace{-10pt}
\end{table*}
\subsubsection{RLS achieves better perceptual scores} We evaluated the reconstruction accuracy of RLS using LPIPS~\cite{zhang2018unreasonable}, PSNR, and MS$-$SSIM~\cite{wang2003multiscale} metrics to compare the reconstructed SR images to the real HR images.
As expected, RLS did not achieve the best pixel-wise reconstruction losses but achieved either the lowest or the second-lowest LPIPS scores on both 32x and 16x upscaling factors. The latter suggested that while the recovered SR image is not exactly the same as the HR image, it is perceptually closer.
It is worth noting that although Pix2Pix produced images with high PSNR and MSSIM scores, it struggled to accurately reconstruct cellular details. This observation highlights the fact that PSNR and SSIM metrics may not be fully appropriate to evaluate the performances of the super-resolution tasks.

\subsubsection{RLS preserves interpretable features}
\begin{figure}[!h]
\vspace{-7pt}
	\centering
	\setlength{\tabcolsep}{0pt}
	\begin{tabular}{DDD@{\hspace{0.1pt}}|@{\hspace{0.1pt}}DDD}
		\multicolumn{3}{c}{\scriptsize DMSO} & \multicolumn{3}{c}{\scriptsize TNF$\alpha$}\\

		\tiny LR & \tiny SR & \tiny GT &
		\tiny LR & \tiny SR & \tiny GT \\

		\MyImmm{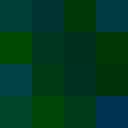} &
		\MyImmm{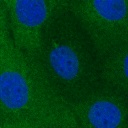} &
		\MyImmm{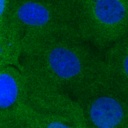} &
		\MyImmm{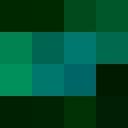} &
		\MyImmm{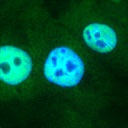} &
		\MyImmm{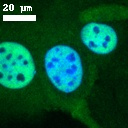} \\
  
		\MyImmm{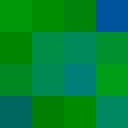} &
		\MyImmm{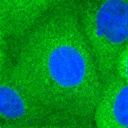} &
		\MyImmm{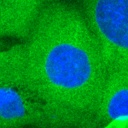} &
		\MyImmm{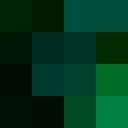} &
		\MyImmm{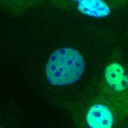} &
		\MyImmm{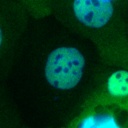} \\

		\MyImmm{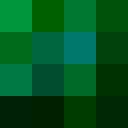} &
		\MyImmm{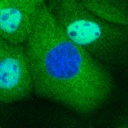} &
		\MyImmm{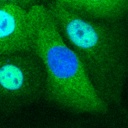} &
		\MyImmm{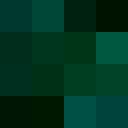} &
		\MyImmm{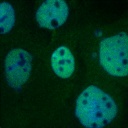} &
		\MyImmm{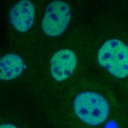} \\

	\end{tabular}
	\caption{
	Visual examples of super-resolving Translocation assay low-resolution images at a 32x upscaling factor
	under negative control (DMSO) and TNF$\alpha$ treatment conditions:
	left: low-resolution image, middle: super-resolution reconstruction, right: ground truth.
	}
	\label{fig:assay_translocation}
 \vspace{-15pt}
\end{figure}
RLS achieves a balance between realism and fidelity when super-resolving microscopy images.
However, as any deep learning method for super-resolution, it cannot generate details it would not have seen during training. Therefore it cannot be used in a context where novel events can be expected. However, it could possibly be used to perform measurements to quantify expected phenotypic changes. This is the case for many assays used in basic research in biology. It is also the case in diagnostics such as parasitemia for instance where the tool must assess if a parasite is present or not. Instead of relying on a blind classification or regression of low-resolution images, these tasks on low-resolution images in such a context could be decomposed in two steps. A first step based on our deep learning approach would consist of reconstructing a high-resolution image, while a second step would use a dedicated handcrafted analysis to quantify a phenotypic feature, making the analysis explicitly interpretable.

To evaluate to what degree the information conveyed by the SR images can be used for such quantitative assays and maybe later for interpretable
diagnostics, we applied it to two assays that allowed straightforward quantification of ``interpretable features'' on HR images but not on LR images.
Here, interpretable features refer to the measurable and explainable phenotypic changes crucial for biological assessment, such as the nucleo-cytoplasmic ratio in response to TNF$\alpha$ treatment and alterations in the Golgi apparatus morphology in reaction to nocodazole. Quantitative analysis was conducted on the SR images to assess the reconstruction of these features, and results were compared against those obtained from HR images as well as baseline methods, including BRGM, PULSE, and an unregularized latent search ('w/o regu.').

\begin{figure}[!h]
    \begin{subfigure}[b]{0.5\linewidth}
        \includegraphics[width=\linewidth]{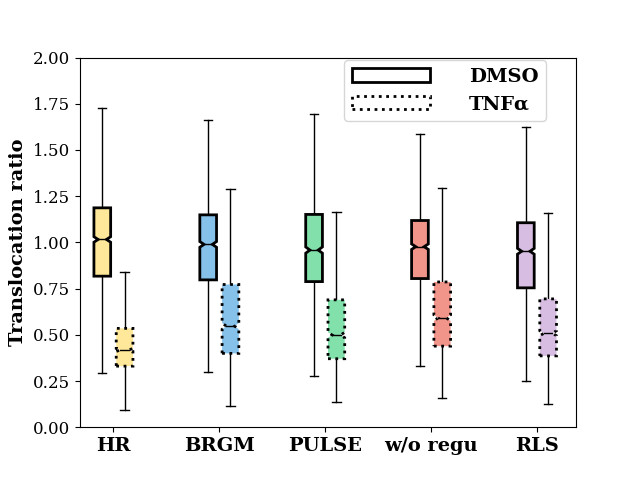}
        \caption{Translocation (32x)}
        \label{fig:measurements-translocation}
    \end{subfigure}
    \hfill
    \hspace{-25pt}
    \begin{subfigure}[b]{0.5\linewidth}
        \includegraphics[width=\linewidth]{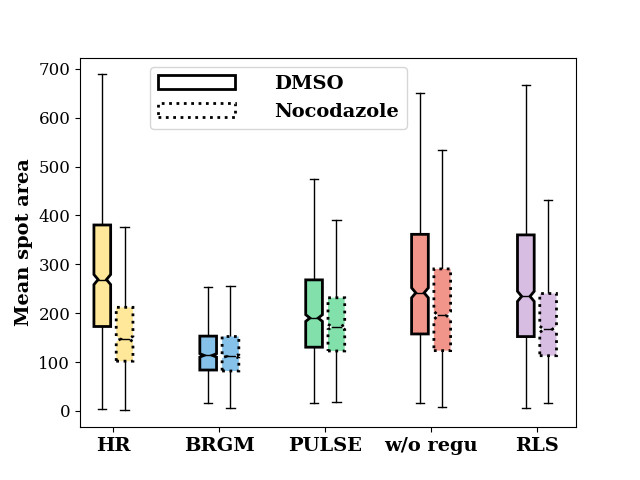}
        \caption{Golgi (16x)}
        \label{fig:measurements-golgi}
    \end{subfigure}
    \caption{
	Making interpretable measurements from low-resolution images: first increasing the resolution by a super-resolution method, then measuring a handcrafted interpretable feature.
	Each pair in the boxplots displays the distribution of handcrafted interpretable measurements, where the solid box represents the negative control (DMSO) and the dotted box signifies the positive controls (TNF$\alpha$ for translocation and Nocodazole for Golgi), across various super-resolution methods including RLS, BRGM, PULSE and ``w/o regu'' alongside with the high-resolution (HR) images for benchmarking.
	\textbf{(a)} Translocation ratio measurement: The y-axis quantifies the translocation ratio, an interpretable metric indicating TNF-induced NF$\kappa$B translocation (green). The translocation ratio can be differentiated between two conditions from real high-resolution images (HR), but also from super-resolution images (SR).
	\textbf{(b)} Mean spot area measurement: The y-axis quantifies the mean spot area, an interpretable metric indicating nocodazole-induced Golgi spreading (green), distinguishable between two conditions from real high-resolution images (HR), but also from super-resolution images (SR) reconstructed by our method.
	}
	\label{fig:measurements}
\end{figure}

The first assay aimed to track the location of the NF-$\kappa$B (nuclear factor kappa B) protein within the cell. Upon
treatment with TNF$\alpha$, a pro-inflammatory cytokine, the protein moves to the nucleus, causing a shift in fluorescence
signal from the cytoplasm to the nuclear area, resulting in bright green nuclei. We observed that the nucleo-cytoplasmic
fluorescence ratio measured on 1000 high-resolution (HR) treated and untreated images, could be replicated when
computed from super-resolved (SR) images (\cref{fig:measurements-translocation}). We also provided some visual examples
of super-resolving the low-resolution images \ie the first step in \cref{fig:assay_translocation}
, which shows that our
method can reconstruct images of wild-type and treated cell images obtained with this common assay.

\begin{figure}[!h]
	\centering
	\setlength{\tabcolsep}{0pt}
	\begin{tabular}{DDD@{\hspace{0.1pt}}|@{\hspace{0.1pt}}DDD}
		\multicolumn{3}{c}{\scriptsize DMSO} & \multicolumn{3}{c}{\scriptsize Nocodazole}\\
		
		\tiny LR & \tiny SR & \tiny GT &
		\tiny LR & \tiny SR & \tiny GT\\
		
		\MyImmm{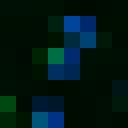} &
		\MyImmm{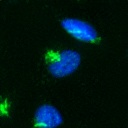} &
		\MyImmm{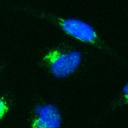} &
  
		\MyImmm{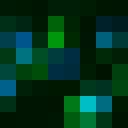} &
		\MyImmm{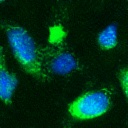} &
		\MyImmm{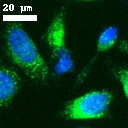} \\
		
		\MyImmm{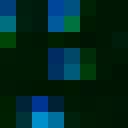} &
		\MyImmm{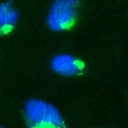} &
		\MyImmm{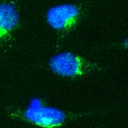} &
		\MyImmm{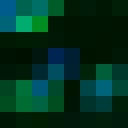} &
		\MyImmm{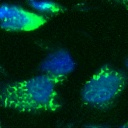} &
		\MyImmm{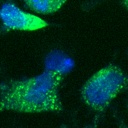} \\
		
		\MyImmm{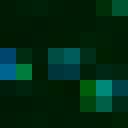} &
		\MyImmm{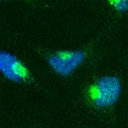} &
		\MyImmm{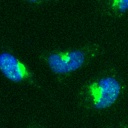} &
		\MyImmm{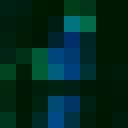} &
		\MyImmm{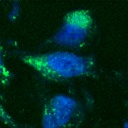} &
		\MyImmm{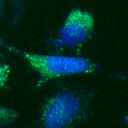} \\
		
	\end{tabular}		
	\caption{
Super-resolution of images from a Golgi assay at a 16x upscaling factor under negative control (DMSO) and Nocodazole
treatment conditions. The left column shows the low-resolution images, the middle column shows the super-resolution reconstructions and the right column shows the ground truth images.
	}
	\label{fig:assay_golgi}
\end{figure}

The second assay we conducted aimed to monitor changes in the morphology of the Golgi apparatus.
When cells are exposed to nocodazole, microtubules disassemble, causing the Golgi, originally located near the center of the cell,
to break up into smaller stacks. \cref{fig:assay_golgi} shows that a standard assay such as the nocodazole-induced Golgi scattering (green)
were reproduced with SR images.
Moreover, as illustrated in \cref{fig:measurements-golgi}, when computed only from SR images,
a straightforward average spot size difference measured on 1000 HR treated and 1000 untreated images could be retrieved.

Furthermore, we utilized cells treated with DMSO as negative controls and cells treated with nocodazole (or TNF in the case of the translocation assay) as positive controls. We developed a classifier trained on high-resolution (HR) images to distinguish between the two phenotypes and evaluated its performance on both HR and super-resolved (SR) images, ensuring that the datasets for GAN training and classifier training were distinct to maintain the integrity of our results. 
Additionally, we trained a separate classifier on low-resolution (LR) images for phenotype differentiation, and its testing was confined to LR images. The results as shown in \cref{tab:accuracy}, indicate that %\added{
the super-resolution operation does not degrade the discriminative signal contained in LR images, the classification accuracy is about the same on SR and LR images. Furthermore, we show evidence that this signal difference can be retrieved when quantifying the same interpretable feature variation present in HR images using handcrafted image analysis (see Fig. \ref{fig:measurements})

It is worth noting that in \cref{fig:measurements} the similarity in performance between the ``w/o regu'' model and RLS during the Translocation assay
reveals an interesting insight.  It suggests that, for this specific assay, the additional
regularization does not significantly change the outcome. However, it’s important to note that this assay was designed to
evaluate strong phenotypic changes where the expected changes are substantial and easy to detect.
In this case, the non-regulated version can also accurately capture the phenotypic changes, since the differences between the
images are large. This is evidenced by the comparative classification accuracies in \cref{tab:accuracy} for the Translocation
assay, where RLS achieves similar performance to the ``w/o regu'' version and the baseline methods.
The advantages of regularization in RLS are more apparent in assays with more subtle phenotypic variations, where detecting
nuanced differences is more challenging. For example, in the Golgi morphology assay, where detecting changes requires
more detailed analysis, RLS's regularization reconstructs realistic images that more faithfully represent biological
structures required for detailed quantitative evaluations.
Furthermore, considering the overall performance, RLS demonstrates consistent and enhanced performance compared to both the baselines
and the non-regularized version, as indicated by the superior classification results in \cref{tab:accuracy}. This suggests that
RLS is a more robust method for super-resolving microscopy images, especially in assays where the phenotypic changes are subtle.
% }
\setlength{\tabcolsep}{2pt}
\begin{table}[!h]
\vspace{-10pt}
	\centering
	\begin{tabular}{|c||cccccc|}
		\hline
		&LR& BRGM& PULSE& w/o Regu.& RLS&HR \\
		\hline
		DMSO vs TNF$\alpha$ (32x) & 85.60 & 79.20 & 85.60 & 85.80 & 87.00 & 97.60 \\
		DMSO vs Nocodazole (16x) & 72.80 & 53.00 & 62.60 & 67.80 & 74.20 & 96.80\\
		\hline
	\end{tabular}
	\caption{
Comparison of classification accuracy for identifying phenotypic changes between negative control (DMSO) and positive control (TNF$\alpha$ and Nocodazole and Nocodazole conditions) in the translocation and Golgi assays respectively, using super-resolved (SR) images and high-resolution (HR) images as a benchmark.}
	\label{tab:accuracy}
	\vspace{-25pt}
\end{table}
Overall, the experiments confirm that the quality of super-resolved images is adequate for further analysis. Our method enables a two-stage process: employing deep learning for the challenging task of super-resolution followed by a handcrafted, interpretable method for the subsequent quantitative measurements.
\begin{figure*}[!h]
	\centering
	\begin{tiny}
		\setlength{\tabcolsep}{0.5pt}
		\begin{tabular}{cDDDDDD}
			& GT & DS
			& \shortstack{\hspace{-7pt}Gaussian Noise \\ ($\sigma=0.05$)}
			& \shortstack{\hspace{-4pt} Gaussian Noise \\ ($\sigma=0.1$)}  
			& \shortstack{Salt\&Pepper \\ ($\sigma=0.05$)}
			& \shortstack{Gaussian Blur \\ ($\sigma=0.5$)}\\
			\multirow{2}{*}{\raisebox{-7mm}{\rotatebox[origin=c]{90}{\MyTiny{\stackanchor{DMSO}{}}}} } &
			\multirow{2}{*}{\MyImmm{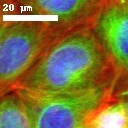}} &
			\MyImmm{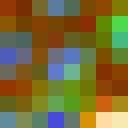}
			&\MyImmm{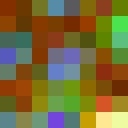}
			& \MyImmm{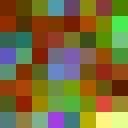}
			& \MyImmm{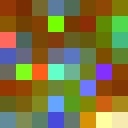}
			& \MyImmm{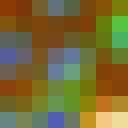}\\
			
			&
			& \MyImmm{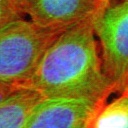}
			& \MyImmm{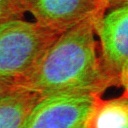}
			& \MyImmm{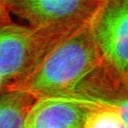}
			& \MyImmm{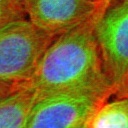}
			& \MyImmm{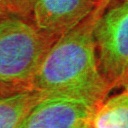}\\

			\multirow{2}{*}{\raisebox{-7mm}{\rotatebox[origin=c]{90}{\MyTiny{\stackanchor{Cytochalasin B}{10$\mu$M}}}}} &
			\multirow{2}{*}{\MyImmm{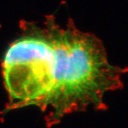}} &
			\MyImmm{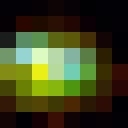}
			&\MyImmm{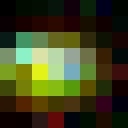}
			& \MyImmm{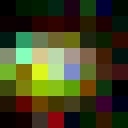}
			& \MyImmm{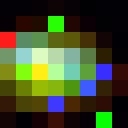}
			& \MyImmm{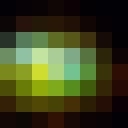}\\
			
			&
			& \MyImmm{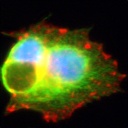}
			& \MyImmm{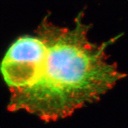}
			& \MyImmm{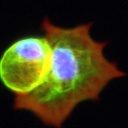}
			& \MyImmm{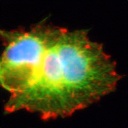}
			& \MyImmm{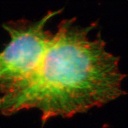}\\
   
			\multirow{2}{*}{\raisebox{-7mm}{\rotatebox[origin=c]{90}{\MyTiny{\stackanchor{Aphidicolin}{10$\mu$M}}}}} &
			\multirow{2}{*}{\MyImmm{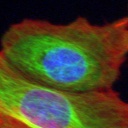}} &
			\MyImmm{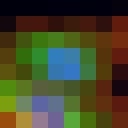}
			&\MyImmm{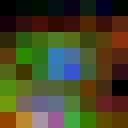}
			& \MyImmm{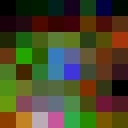}
			& \MyImmm{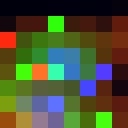}
			& \MyImmm{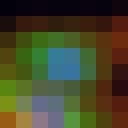}\\
			
			&
			& \MyImmm{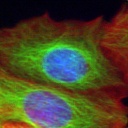}
			& \MyImmm{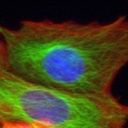}
			& \MyImmm{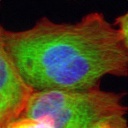}
			& \MyImmm{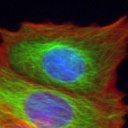}
			& \MyImmm{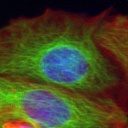}\\
		\end{tabular}
	\end{tiny}
	\caption{
	Evaluation of RLS performance under various degradation conditions:
	images downscaled by bicubic method are further altered with Gaussian noise, Salt and Pepper noise, and Gaussian blur
	to assess the stability of the proposed method across a range of image perturbations (at a 16x upscaling factor).
	}
	\label{fig:robustness}
\end{figure*}

\subsection{Robustness}
\label{sec: robustness}
In opposition to supervised methods that are sensitive to the input image domain, this approach is not restricted to a particular degradation operator that is used during training. To evaluate this aspect we applied additional degradation
operators such as Gaussian noise, Salt and Pepper, and Gaussian blur to a bicubic downscaled image DS before reconstruction.
As depicted in \cref{fig:robustness}, the reconstruction closely matches the DS image.
This result validates our choice of using the bicubic downscaling operator during training instead of more complicated specific degradation.

\begin{figure}[!h]
  \centering
  \begin{minipage}[b]{0.75\linewidth}
    \hspace{-23pt}
        \setlength{\tabcolsep}{0pt}
	\begin{scriptsize}
		\begin{tabular}{FFFFFFF}
		& LR & w/o Regu. & w/o $p_{w}$ & w/o $p_\text{cross}$ & RLS & GT\\
            \raisebox{7mm}{\rotatebox[origin=c]{90}{\MyTiny{\stackanchor{DMSO}{}}}} &
            \MyImn{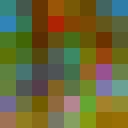} &
            \MyImn{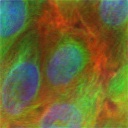} &
			\MyImn{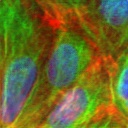} &
			\MyImn{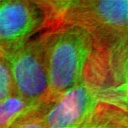} &
			\MyImn{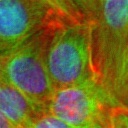} &
			\MyImn{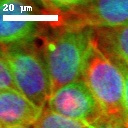}\\

            \raisebox{7mm}{\rotatebox[origin=c]{90}{\MyTiny{\stackanchor{Brefeldin A}{0.01$\mu$M}}}} &
            \MyImn{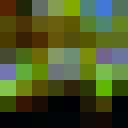} &
            \MyImn{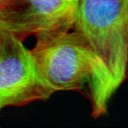} &
			\MyImn{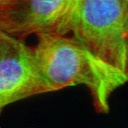} &
			\MyImn{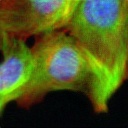} &
			\MyImn{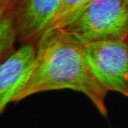} &
			\MyImn{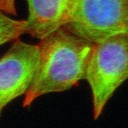}\\

            \raisebox{7mm}{\rotatebox[origin=c]{90}{\MyTiny{\stackanchor{Cytochalasin D}{10$\mu$M}}}} &
            \MyImn{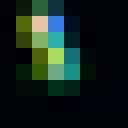} &
            \MyImn{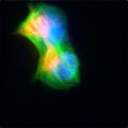} &
			\MyImn{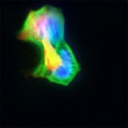} &
			\MyImn{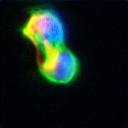} &
			\MyImn{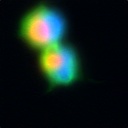} &
			\MyImn{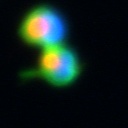}\\

            \raisebox{7mm}{\rotatebox[origin=c]{90}{\MyTiny{\stackanchor{Nocodazole}{3$\mu$M}}}} &
            \MyImn{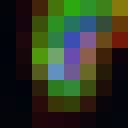} &
			\MyImn{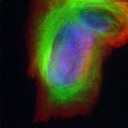} &
			\MyImn{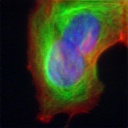} &
			\MyImn{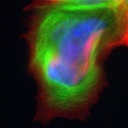} &
			\MyImn{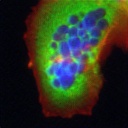} &
			\MyImn{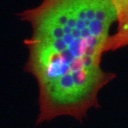}\\
            
            \raisebox{7mm}{\rotatebox[origin=c]{90}{\MyTiny{\stackanchor{cytochalasin B}{30$\mu$M}}}} &
            \MyImn{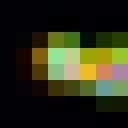} &
            \MyImn{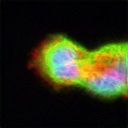} &
			\MyImn{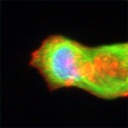} &
			\MyImn{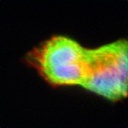} &
			\MyImn{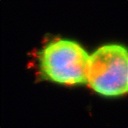} &
			\MyImn{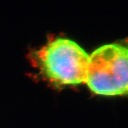}\\
		\end{tabular}
	\end{scriptsize}
        \label{fig:subfig-ablation-a}
    \end{minipage}
    \hfill
  \begin{minipage}{0.22\linewidth}
  \vspace{10pt}
\centering
\hspace{-30pt}
     \setlength{\tabcolsep}{0.1pt}
	\begin{tabular}{@{}l|ll@{}}
		\toprule
		% Parameter & Symbol & Value & Unit \\
		Variant & FID$\downarrow$ & PSNR$\uparrow$\\
		\hline
		w/o Regu. & 14.69 & 20.56\\
		w/o $p_{w}$ & 7.96 & 20.21\\
        w/o $p_\text{cross}$ & 8.99 & 21.24\\
		RLS & 6.98 & 19.18\\[0.05cm]
		\bottomrule
	\end{tabular}
        \label{fig:subfig-ablation-b}
    \end{minipage}
    \caption{
Ablation study showcasing the impact of regularization components on RLS performance,
	with qualitative results in the left column and quantitative results in the right column.
	The variants include: ``w/o Regu.'' (searching the latent space without any regularization),
	``w/o $p_{w}$'' (the image prior does not include the prior term $p_{w}$),
	``w/o $p_\text{cross}$'' (the image prior does not include the prior term $p_\text{cross}$), and
	``RLS'' (the full RLS model) (at a 16x upscaling factor).
	}
    \label{fig:ablation}
\end{figure}

\subsection{Ablation study}

\cref{fig:ablation} demonstrates ablation experiments highlighting the impact of the components of our image prior. First, ``w/o Regu.'' searches the latent space without any regularization for the image that, once downscaled matches the LR image.
The second variant is denoted 
``w/o $\prior_{w}$'', \ie, the image prior does not include the prior term $\prior_{w}$. Similarly, ``w/o $\prior_\text{cross}$'' refer to the suppression of 
$\prior_{cross}$.

To evaluate the three variants, we use the same set of parameters described in \cref{experimental_setting}. One can see that searching the latent space without any regularization produces images that do not necessarily belong to the original image manifold and therefore do not appear realistic. It also tends to generate images that are accurately downscaled to the LR image but at the cost of generating distorted images when $\prior_\text{cross}$ or $\prior_{w}$ is discarded.
This implies that both priors play an important role in generating realistic details.

\subsection{Uncertainty.}
\begin{figure*}[!h]
 	\centering
 	\setlength{\tabcolsep}{0.5pt}
 	\begin{tiny}
 		\begin{tabular}{CCCCCCC}
 			LR & GT & SR$_{1}$ & SR$_{2}$ &SR$_{3}$ & SR$_{4}$ & SR$_{5}$\\
 			
 			\MyImm{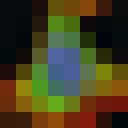} &
 			\MyImm{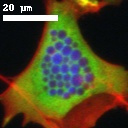} &
 			\MyImm{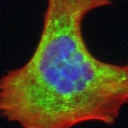} &
			\MyImm{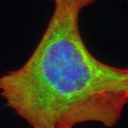} &
			\MyImm{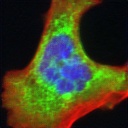} &
			\MyImm{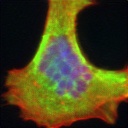} &
			\MyImm{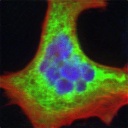}\\

 			\MyImm{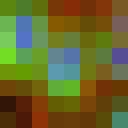} &
 			\MyImm{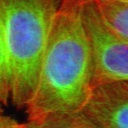} &
			\MyImm{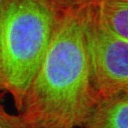} &
			\MyImm{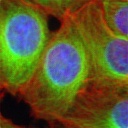} &
			\MyImm{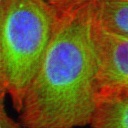} &
			\MyImm{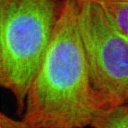} &
			\MyImm{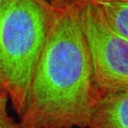}\\

			\MyImm{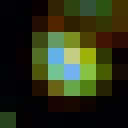} &
 			\MyImm{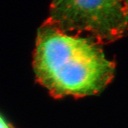} &
			\MyImm{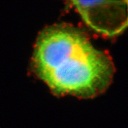} &
			\MyImm{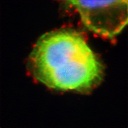} &
			\MyImm{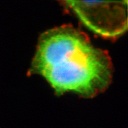} &
			\MyImm{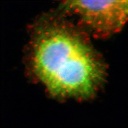} &
			\MyImm{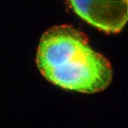}\\
 		\end{tabular}
 	\end{tiny}
 	\caption{
Visualizing the ill-posed nature of the super-resolution task and the uncertainty associated with SR reconstruction. Five distinct SR images (SR$_{1}$ to SR$_{5}$) are generated for each LR image by sampling five different latent codes from the latent space, alongside the ground truth (GT) (at a 16x upscaling factor).}
 	\label{fig:uncertainty}
 \end{figure*}

An important challenge of the super-resolution task is that it is an ill-posed problem. Although we can improve this aspect by using an image prior constraint, several closely related highly resolved images could still be consistent with a single LR image.
To generate $n$ realistic SR images,
we sample $n$ latent codes denoted as $\w_1, \w_2, \dots, \w_n$. We assume that their distribution follows a Gaussian distribution $\normal(\mu, \sigma)$, where the parameter $\sigma$ follows an inverse gamma distribution.
Using the Bayes rule, we estimate the distribution's parameters by:
$$\max_{\mu, \sigma} \left[\sum_{i=1}^{n} \log p(\w_i|y) - \frac{n}{2}\log\sigma^2 - \frac{n}{2}\log(2\pi) - \frac{1}{2\sigma^2}\sum_{i=1}^{n} (\w_i-\mu)^2 + \log p(\sigma)\right]$$
Here, the first term is the log-likelihood of the posterior distribution of $p(\w|\y)$, which is defined in \cref{equation loss}, the second term is the regularization term, which penalizes large values of $\sigma$ and $\log p(\sigma)$ is the log prior distribution of $\sigma$. \cref{fig:uncertainty} displays multiple solutions for a given LR image we can obtain with our approach. Sampling close variations of HR images from a single LR input can be used to enhance the robustness of the estimation of an image-based quantitative feature by reducing the effects of noise or artifacts that may exist in the input LR image.

\section{Conclusion}
\label{sec:conclusion}
In this paper, we propose a robust super-resolution method based on regularized latent search within a pre-trained StyleGAN.
It does not require coupled image pairs for training and constraints the super-resolution task to a given image prior to offer a trade-off between fidelity and realism of the SR reconstruction. Furthermore, we show that such a method could
be used to split analyses, such as the classification of low-resolution images, into a reconstruction of super-resolution
images performed by deep learning and a simple dedicated handcrafted analysis
of an interpretable feature.

The latter could be used for instance for rapid diagnostic
based on smartphone directly
available on the field. In this case, a dataset of low-resolution images of slides acquired from fast and/or cheap solutions available on the field could be coupled with the acquisition of the same slides using high-end expensive high-resolution microscopes with limited access. In this application, high-resolution images could then be reconstructed directly on the field to perform an explainable diagnostic such as a parasite count.

\section{Limitations and future work}
\label{sec:limitations}

One common challenge with deep learning-based methods, especially those using generative priors like in our study, is their limited generalizability to unfamiliar and unseen data. Our approach, which utilizes StyleGAN for unsupervised learning within a specific domain, may not encompass the full diversity encountered in real-world scenarios. To enhance the model's generalizability, strategies such as incorporating data augmentation to introduce training data variability, or applying transfer learning for domain adaptation, could be beneficial.\\
Our current findings serve as an initial validation of our method's capabilities. However, we acknowledge the necessity of further evaluations using more varied datasets, encompassing a broader range of imaging techniques and sample preparation methods. The absence of such diverse experimental data in our current research is due to the specific requirement of our study to explore super-resolution at high upscaling factors, like 16x and 32x. These factors, which are crucial to our application's needs, are rarely addressed in the available literature, and to our knowledge, datasets with such extreme upscaling factors are not yet available. Existing datasets generally focus on more moderate upscaling factors, such as 2x or 4x, which do not meet the demands of our research that targets significantly higher levels of resolution enhancement.
We are committed to extending our validation to include real-world imaging conditions as soon as datasets meeting our high upscaling factor requirements become available. This will enable a more comprehensive assessment of our method's applicability and performance in practical scenarios.

\section{Acknowledgments}
\label{sec:acknowledgments}
We thank GENCI for the access to the HPC resources of IDRIS under the allocation 2020-AD011011495.

\section{Competing interests}
The authors declare no competing interests.

\section{Author contributions}
MG proposed and implemented the RLS method. AG hypothesized that RLS reconstructs interpretable features and suggested experiments to validate it. MG run the experiments. 
MG and AG wrote the manuscript.

\section{Funding statement}
This work was supported by the ANR VISUALPSEUDOTIME, ANR–10–LABX–54 MEMOLIFE and ANR–10 IDEX 0001–02 PSL* Université Paris.

\section{Data availability}
We used the BBBC021 image set available from the Broad Bioimage Benchmark Collection (https://bbbc.broadinstitute.org/).

\bibliographystyle{unsrt}
\bibliography{ms}
\end{document}

%% file: commands.tex
\newcommand\func[1]{\mathsf{#1}} % functions
\newcommand\vect[1]{\mathbf{#1}}
\def \real {\mathbb{R}}
\def \wspace {\mathcal{W}}
\def \zspace {\mathcal{Z}}
\def \normal {\mathcal{N}}
\def \sphere {\mathcal{S}}
\def \w {\vect{w}}

\def \z {\vect{z}}
\def \x {\vect{x}}
\def \y {\vect{y}}
\def \F {\func{F}}
\def \G {\func{G}}
\def \D {\func{D}}
\def \dim {d}
\def \layer{L}
\def \n {n}
\def \zero {\vect{0}}
\def \eye {\vect{I}}
\def \Pg {p_{\G}}
\def \Pf {p_{\F}}
\def \deltaa {\delta}
\def \Pdelta {p_{\deltaa}}
\def \Pw {p_{\w}}
\def \prior {\mathcal{P}}
\def\onedot{. }
 
\def\ie{\emph{i.e}\onedot} 
\newcommand{\etal}{\textit{et al.}}